\documentclass{article}

\usepackage[cp1252]{inputenc}

\usepackage[english]{babel}

\usepackage{amssymb}
\usepackage{amsmath}

\begin{document}

\selectlanguage{english}

\noindent \textbf{\large On the physical field generated by rotating masses}

\noindent \textbf{\large in Poincar\'{e}-gauge theory of gravity}

\textbf{}

\noindent\textbf{\large B.N. Frolov}

\noindent
Department of Physics for Nature Sciences, Moscow State Pedagogical University,\\
Krasnoprudnaya 14, Moscow 107140, Russian Federation. \\
E-mail: frolovbn@mail.ru; frolovbn@orc.ru

\textbf{}

\noindent
 It is shown that the gauge field in Poincar\'{e}-gauge theory of gravity consists in two parts: the
translational\textit{ }gauge field ( $t$ -field), which is generated by the energy-momentum current of external
fields, and the rotational\textit{ }gauge field ( $r$ -field), which is generated by the sum of the angular and
spin momentum currents of external fields. In connection with this the physical field generating by rotating
masses should exist.

\textbf{}

\noindent\textbf{\large 1. Introduction}

\textbf{}

\noindent The problem of a great interest consists in a possible existing of still unknown physical field that can
be generated by rotating masses as a source. In particular, such possibility follows from the Theorem on the
sources of the gauge fields [1--4], which states that the gauge field is generated by the Noether invariant
corresponding to the Lie group that introduces this gauge field by the localization procedure. In this sense the
gauge field in Poincar\'{e}-gauge theory of gravity (PGTG) should be generated not only by the energy--momentum
tensor, but also by the sum of angular and spin momentum currents as the source.

Here we shall discuss the following problems:

\begin{itemize}
\item What are the true gauge potentials in PGTG?

\item What are the source currents of the field equations in PGTG?

\item What fields can be generated by rotating masses in PGTG?
\end{itemize}

The corresponding field equations of PGTG were derived in [3,4]  as the consequence of the general gauge field
theory for the groups connecting with space-time transformations. We have shown [5] that under the localization
procedure the Lorenz subgroup of the Poincar\'{e} group introduces the \textit{rotational }gauge field  $A_{a}^{m}
\, $ ( $r$ -field), which is generated by the sum of the angular and spin momentum currents of external fields.
The subgroup of translations introduces the \textit{translational }gauge field  $A_{a}^{k} \, $ ( $t$ -field),
which is generated by the energy-momentum current of external fields. These field equations are equivalent to the
equations of PGGT in usual form, derived by the variation of the Lagrangian with respect to the tetrads  $h^{a}\!
_{\mu } \, $  and connections  $A^{m}\! _{\mu } \, \, =\, \, A^{m} _{a} \, h^{a}\! _{\mu } \, $ , the tetrads and
also curvature and torsion being constructed with the help of both the  $t$ - and  $r$ -fields.

\textbf{}

\noindent \textbf{\large 2. Noether theorem and the principle of local invariance}
\setcounter{section}{2}\setcounter{equation}{0}

\textbf{}

We start with the flat Minkowski space  ${\rm M}_{{\rm 4}} $  with the Cartesian coordinates  $x^{a} \, $  ( $a\,
=\, 1,\, 2,\, 3,\, 4$ ) and the metric  $g_{ab} \, =\, \breve{g}(\vec{e}_{a} \, ,\, \vec{e}_{b} )\, =\, {\rm
diag}\, {\rm (1,}\, {\rm 1,}\, {\rm 1,}\, {\rm -1)}\, $ with the basis  $\vec{e}_{a} \, =\, \partial _{a} \, =\,
{\partial \mathord{\left/ {\vphantom {\partial  \partial x^{a} }} \right. \kern-\nulldelimiterspace} \partial
x^{a} } $ . The fundamental group of  ${\rm M}_{{\rm 4}} $  is Poincar\'{e} group  ${\rm P}_{{\rm 4}} \, $
(inhomogeneous Lorentz group),

\[
\begin{array}{l}
{\delta x^{a} \, = \, \omega ^{m} I_{m}\!^{a}\!_{b} \, x^{b} \, \, +\, \, a^{a} \, \, =\, \, \omega ^{z} X_{z}^{a}
\, \, =\, \, -\omega ^{z} \hat{M}_{z} \, x^{k} \, ,\, \, \, \, \, \, \, \, \, \, \{ \omega ^{z} \} \, \, =\, \, \{
\omega ^{m} ,\, \, a^{k} \} \, ,\, } \\ {\hat{M}_{z} \, \, =\, \, \{ \hat{M}_{m} \, ,\, \, \hat{M}_{k} \} \, ,\,
\, \, \, \, \, \, \, \, \, \hat{M}_{m} \, \, =\, \, -I_{m}\!^{a}\!_{b} \, x^{b} \frac{\partial }{\partial x^{a} }
\, ,\, \, \, \, \, \, \, \, \, \, \hat{M}_{k} \, \, =\, \, P_{k} \, \, =\, \, -\frac{\partial }{\partial x^{k} }
\, .}
\end{array}
\]
Here we have introduced the abbreviations for the rotations and translations,
\[
X_{z}^{a} \, \, =\, \, \{ X_{m}^{a} ,\, \, X_{k}^{a} \} \, ,\, \, \, \, \, \, \, \, \, X_{m}^{a} \, \, =\, \,
I_{m}\!^{a}\!_{b} \, x^{b} \, ,\, \, \, \, \, \, \, \, \, X_{k}^{a} \, \, =\, \, \delta _{k}^{a} \, .
\]

Let us introduce the curvilinear system of coordinate  $x^{\mu } (x^{a} )\, $ on  ${\rm M}_{{\rm 4}} $ :
\[
\begin{array}{l}
{dx^{a} \, \, =\, \, \hat{h}^{a}\! _{\mu } \, dx^{\mu } \, ,\, \, \, \, \, \, \, \, \, \, \hat{h}^{a}\! _{\mu } \,
\, = \, \, \vec{e}_{\mu } (x^{a} )\, ,\, \, \, \, \, \, \, \, \, \, \vec{e}_{\mu } \, \, =\, \, \partial _{\mu }
\, \, = \, \, {\partial \mathord{\left/ {\vphantom {\partial  \partial x^{\mu } \, ,}} \right.
\kern-\nulldelimiterspace}
\partial x^{\mu } \, ,} \, \, \, \, \, \, \, \, \, \, ds^{2} \, \, =\, \, g_{\mu \nu } \, dx^{\mu } \, dx^{\nu } \, ,} \\
{\hat{g}_{\mu \nu } \, \, =\, \, g_{ab} \, \hat{h}^{a}\! _{\mu } \, \hat{h}^{b}\! _{\nu } \, ,\, \, \, \, \, \, \,
\, \, \, \hat{g}\, \, =\, \, {\rm det}\, {\rm (}\hat{g}_{\mu \nu } {\rm )}\, \, {\rm =}\, \, {\rm det}\, (g_{ab}
{\rm )}\, \hat{h}^{2} \, ,\, \, \, \, \, \, \, \, \, \, \hat{h}\, \, =\, \, {\rm det}\, (\hat{h}^{a}\! _{\mu } )\,
\, =\, \, \sqrt{\left|\hat{g}\right|} \, .\, }
\end{array}
\]

Action integral is invariant under the Poincar\'{e} group  ${\rm P}_{{\rm 4}} \, $ ,
\[
J\, \, =\, \, \int _{\Omega }(dx)\, \sqrt{\left|\hat{g}\right|} \, L(Q^{A} ,\, P_{k} Q^{A} )\, ,\, \, \, \, \, \,
\, \, \, P_{k} \, \, =\, \, -\hat{h}^{\mu } _{k} \, \partial _{\mu } \, ,\, \, \, \, \, \, \, \, \, \, {\cal L}\,
\, =\, \, \sqrt{\left|\hat{g}\right|} \,  L\, \, =\, \, \hat{h}\, L\, .
\]
The first Noether theorem in a curvilinear system of coordinate yields the following,
\[
0\, \, =\, \, \int _{\Omega }(dx)\, \left[\frac{\delta {\cal  L}}{\delta Q^{A} } \, \bar{\delta }Q^{A} \, \, +\,
\,
\partial _{\mu } \left({\cal  L}\, \hat{h}^{\mu }\! _{k} \, \delta x^{k} \, \, -\, \, \hat{h}^{\mu }\! _{k}
\frac{\partial {\cal  L}}{\partial P_{k} Q^{A} } \, \bar{\delta }Q^{A} \right)\right] \,.
\]

Here  $\bar{\delta }$  denotes the variation of the form of the field. For example, for the field  $Q^{A} $ we
have,
\[
\bar{\delta }Q^{A} \, \, =\, \, \delta Q^{A} \, \, -\, \, \delta x^{k} \, \partial _{\mu } \, Q^{A} \,.
\]
The field equation of the field  $Q^{A} $  is fulfilled,
\[
0\, \, =\, \, \frac{\delta {\cal  L}}{\delta Q^{A} } \, \, =\, \, \frac{\partial {\cal  L}}{\partial Q^{A} } \, \,
+\, \, \partial _{\mu } \left(\hat{h}^{\mu }\! _{k} \, \frac{\partial {\cal  L}}{\partial P_{k} Q^{A} } \right)\,
,
\]
The result of Noether theorem can be represented as follows,
\[
0\, \, =\, \, \int _{\Omega }(dx)\, \hat{h}\hat{h}^{\mu }\! _{k} \, \partial _{\mu } \left(a^{l} \, t^{k}\! _{l}
\, +\, \, \omega ^{m} \, M^{k}\! _{m} \right) \, ,
\]
where the following expressions for the energy-momentum  $t^{k}\! _{l} \, $ and the full momentum  $M^{k}\! _{m} $
(angular momentum plus spin momentum  $J^{k}\! _{m} $  ) tensors are introduced,
\[
t^{k}\! _{l} \, \, =\, \, L\delta _{l}^{k} \, \, -\, \, \frac{\partial L}{\partial P_{k} Q^{A} } P_{l} Q^{A} \, ,
\, \, \, \, \, \, \, \, \, M^{k}\! _{m} \, \, =\, \, J^{k}\! _{m} \, \, +\, \, I_{m}\!^{l}\!_{b} \, x^{b} t^{k}\!
_{l} \, ,\, \, \, \, \, \, \, \, \, \, J^{k}\! _{m} \, \, =\, \, -\frac{\partial L}{\partial P_{k} Q^{A} } \,
I_{m}\!^{A}\!_{B} \, Q^{A} \, .
\]
Noether Theorem yields the conservation laws for the energy-momentum and full momentum,
\[P_{k} \, t^{k}\! _{l} \, \, =\, \, 0\, ,\, \, \, \, \, \, \, \, \, \, P_{k} \, M^{k}\! _{m} \, \, =\, \, 0\, .\]

Now we shall \textit{localize} the Poincar\'{e} group  ${\rm P}_{{\rm 4}} \, $ , the parameters of which become
arbitrary functions of coordinates on  ${\rm M}_{{\rm 4}} $ . The theory is based on four Postulates.

\noindent \textbf{Postulate 1} \textit{(The principle of local invariance).} The action integral
\begin{equation} \label{GrindEQ__2_1_}
J\, \, =\, \, \int _{\Omega }(dx)\, {\cal  L}(Q^{A} ,\, P_{k} Q^{A} ;\, \, A_{a}^{R} ,\, P_{k} A_{a}^{R} )\, ,\,
\end{equation}
where the Lagrangian density  ${\cal  L}$  describes the field  $Q^{A} $ , interaction of this field with the
additi-onal gauge field  $A_{a}^{R} $  and also the free gauge field  $A_{a}^{R} $ , is invariant under the action
of the localized group  ${\rm P}_{{\rm 4}} (x)\, $ , the gauge field being transformed as follows,
\begin{equation}\label{GrindEQ__2_2_}
\delta A_{a}^{R} \, \, =\, \, U^{R}_{za} \, \omega ^{z} \, \, +\, \, S^{R\mu} _{za} \,
\partial _{\mu } \omega ^{z} \, ,
\end{equation}
where  $U\, $ and  $S$  are unknown matrices.

\noindent \textbf{Postulate 2} \textit{(The principle of stationary action)}:
\begin{equation} \label{GrindEQ__2_3_} \frac{\delta {\cal  L}}{\delta Q^{A} } \, \, =\, \, 0\, ,\, \, \, \, \, \,
\, \, \, \, \, \, \, \, \, \, \frac{\delta {\cal  L}}{\delta A_{a}^{R} } \, \, =\, \, 0\, .
\end{equation}

\noindent \textbf{Postulate 3} \textit{(An independent existence of a free gauge field)}. The full Lagrangian
density ${\cal L}$  of the physical system has the following structure,
\begin{equation} \label{GrindEQ__2_4_} {\cal  L}\, \, =\, \, {\cal  L}_{0} \, \, +\, \, {\cal  L}_{Q} \, ,\, \, \, \,
\, \, \, \, {\cal  L}_{0} \, \, {\rm =}\, \, {\cal  L}_{0} (A_{a}^{R} ,\, \, P_{k} A_{a}^{R} )\, ,\, \, \, \, \, \,
\, \, \, \, \frac{\partial {\cal  L}_{0} }{\partial Q^{A} } \, \, =\, \, 0\, ,\, \, \, \, \, \, \, \,
\frac{\partial {\cal  L}_{0} }{\partial P_{k} Q^{A} } \, \, =\, \, 0\, .
\end{equation}

\noindent \textbf{Postulate 4} \textit{(The principle of minimality of gauge interaction)}:
\begin{equation} \label{GrindEQ__2_5_}
\frac{\partial {\cal  L}_{Q} }{\partial P_{k} A_{a}^{R} } \, \, =\, \, 0\, .
\end{equation}

The second Noether theorem for the Lagrangian density \eqref{GrindEQ__2_1_} and the group  ${\rm P}_{{\rm 4}}
(x)\, $ yields the following,

\begin{equation} \label{GrindEQ__2_6_}
\begin{array}{l}
{0\, \, =\, \, \int _{\Omega }(dx)\, \left[\frac{\delta {\cal  L}}{\delta Q^{A} } \, \bar{\delta }Q^{A} \, \, +\,
\, \frac{\delta {\cal  L}} {\delta A_{a}^{R} } \, \bar{\delta }A_{a}^{R} \right] \, \, } \\ {+\, \, \int _{\Omega
}(dx) \, \partial _{\mu } \left({\cal  L}\, \hat{h}^{\mu }\! _{k} \, \delta x^{k} \, \, -\, \, \hat{h}^{\mu }\!
_{k} \frac{\partial {\cal  L}} {\partial P_{k} Q^{A} } \, \bar{\delta }Q^{A} \, \, +\, \, \hat{h}^{\mu }\! _{k}
\frac{\partial {\cal  L}} {\partial P_{k} A_{a}^{R} } \, \bar{\delta }A_{a}^{R} \right).}
\end{array}
\end{equation}
Here in  $\delta x^{k} ,\, \, \bar{\delta }Q^{A} ,\, \, \bar{\delta }A_{a}^{R} $ we have arbitrary functions
$\omega ^{z} (x),\, \, \partial _{\mu } \omega ^{z} (x),\, \, \partial _{\mu } \partial _{\nu } \omega ^{z} (x)$ ,
coefficients before them being equal to zero,
\begin{equation} \label{GrindEQ__2_7_}
\begin{array}{l}
{\partial _{\mu } \left(\hat{h}\hat{h}^{\mu }\! _{k} \, \Theta _{z}^{k} \right)\, \, +\, \, \left(U_{za}^{R} \, \,
+\, \, X_{z}^{l} \, P_{k} \, A_{a}^{R} \right) \frac{\delta {\cal  L}}{\delta A_{a}^{R} } \, \, =\, \, 0\, ,} \\
{\hat{h}\hat{h}^{\mu }\! _{k} \, \Theta _{z}^{k} \, \, +\, \, \partial _{\nu } \, {\rm M}^{\nu \mu }\! _{z} \, \,
+\, \, S_{za}^{R\mu } \frac{\delta {\cal  L}}{\delta A_{a}^{R} } \, \, =\, \, 0\, ,\, \, \, \, \, \, \, \, \, \,
\, \, \, \, {\cal M}^{(\nu \mu )}\! _{z} \, \, =\, \, 0\, .}
\end{array}
\end{equation}
where the following notations are introduced,
\begin{equation} \label{GrindEQ__2_8_}
\begin{array}{l}
{\hat{h}\, \Theta _{z}^{k} \, \, =\, \, {\cal  L}\, X_{z}^{k} \, \, -\, \, \left(I_{m}\!^{A}\!_{B} \, Q^{B} \, \,
+\, \, X_{z}^{l} \, P_{l} \, Q^{A} \right)\frac{\partial {\cal  L}}{\partial P_{k} Q^{A} } \, \, -\, \,
\left(U_{za}^{R}
\, \, +\, \, X_{z}^{l} \, P_{k} \, A_{a}^{R} \right)\frac{\partial {\cal  L}}{\partial P_{k} A_{a}^{R} } \, ,} \\
{{\cal M}^{\nu \mu }\! _{z} \, \, =\, \, \hat{h}^{\nu }\! _{k} \, \frac{\partial {\cal  L}}{\partial P_{k}
A_{a}^{R} } \, S_{za}^{R\mu } \, .}
\end{array}
\end{equation}
If the equations \eqref{GrindEQ__2_3_} for the gauge field are valid, then the equations \eqref{GrindEQ__2_7_} are
simplified,
\begin{equation} \label{GrindEQ__2_9_}
\partial _{\mu } \left(\hat{h}\hat{h}^{\mu }\! _{k} \,
\Theta _{z}^{k} \right)\, \, ,\, \, \, \, \, \, \, \, \, \, \hat{h}\hat{h}^{\mu }\! _{k} \, \Theta _{z}^{k} \, \,
+\, \, \partial _{\nu } \, {\cal M}^{\nu \mu }\! _{z} \, \, =\, \, 0\, ,\, \, \, \, \, \, \, \, \, \, {\cal
M}^{(\nu \mu )}\! _{z} \, \, =\, \, 0\, .
\end{equation}

\textbf{}

\noindent \textbf{\large 3. Structure of the Lagrangian densities  ${\cal  L}_{Q} $  and  ${\cal  L}_{0} $  }
\setcounter{section}{3}\setcounter{equation}{0}

\textbf{}

\noindent We introduce the differential operator  $M_{R} $ ,
\begin{equation} \label{GrindEQ__3_1_}
M_{R} \, \, =\, \, \{ M_{m} ,\, \hat{M}_{k} \} \, ,\, \, \, \, \, \, \, \, \, \, M_{m} \, \, =\, \,
\hat{M}_{m} \, \, +\, \, I_{m} \, ,\, \, \, \, \, \, \, \, \, \, \hat{M}_{k} \, \, =\, \, P_{k} \, .
\end{equation}
and represent the gauge field in two components,  $A_{a}^{R} \, \, =\, \, \{ A_{a}^{m} ,\, A_{a}^{k} \} \, $ ,
where  $A_{a}^{k} $  describes the translational part of the gauge field ( $t$ -field), and  $A_{a}^{m} $
describes the rotational part of the gauge field ( $r$ -field).

\noindent \textbf{Theorem 1} (B.N. Frolov, 1999, 2003). The gauge field  $A_{a}^{R} $  exists with
transformational structure of Postulate 1 under the localized Poincar\'{e} group  ${\rm P}_{{\rm 4}} (x)\, $ , and
such the matrix functions  $Z$ ,  $U$  and  $S$  of the gauge field exist that the Lagrangian density,
\begin{equation} \label{GrindEQ__3_2_}
{\cal  L}_{Q} \, \, =\, \, h\, L_{Q} (Q^{A} ,\, D_{a} Q^{A} )\, ,\, \, \, \, \, \, \, \, \, \, \, \, \, \, \,
h\, =\, \, Z\, \hat{h}\, ,
\end{equation}
satisfies to Postulate 1,  ${\cal  L}_{Q} $  being constructed from  $L_{Q} (Q^{A} ,\, P_{a} Q^{A} )\, $  by
exchanging the operator  $P_{a} $  with the operator of the gauge derivative,
\begin{equation} \label{GrindEQ__3_3_}
 D_{a} \, \, =\, \, -A_{a}^{R} \, M_{R} \, .
\end{equation}
Also the following representation for the gauge  $t$ -field is valid,
\begin{equation} \label{GrindEQ__3_4_} A_{a}^{k} \, \, =\, \, D_{a} \, x^{k} \, . \end{equation}
The \textbf{\textit{proof}} of this Theorem has been performed in [3,4] and consists in proving three
Prepositions.

\noindent \textbf{\textit{Preposition 1.1}}. With the help of \eqref{GrindEQ__3_1_} the gauge derivative
\eqref{GrindEQ__3_3_} can be represented as follows,
\begin{equation} \label{GrindEQ__3_5_}
\begin{array}{l}
{D_{a} Q^{A} \, \, =\, \, h^{\mu }\! _{a} \, \partial _{\mu } Q^{A} \, \, -\, \, A_{a}^{m} \, I_{m}\!^{A}\!_{B} \,
Q^{B} \, \, =\, \, h^{\mu }\! _{a} \, D_{\mu } Q^{A} \, ,} \\ {D_{\mu } Q^{A} \, \, =\, \,
\partial _{\mu } Q^{A} \, \, -\, \, A^{m}\! _{\mu } \, I_{m}\!^{A}\!_{B} \, Q^{B} \, ,}
\end{array}
\end{equation}
where new quantities are introduced,
\begin{equation} \label{GrindEQ__3_6_}
\begin{array}{l}
{Y_{a}^{k} \, \, =\, \, A_{a}^{k} \, \, + \, \, A_{a}^{m} \, X_{m}^{k} \, \, =\, \, A_{a}^{R} \, X_{R}^{k} \, ,\,
\, \, \, \, \, \, \, \, \, h^{\mu }\! _{a} \, \, =\, \, \hat{h}^{\mu }\! _{k} \, Y_{a}^{k} \, \, ,} \\ {h^{a}\!
_{\mu } \, \, = \, \, Z_{k}^{a} \, \hat{h}^{k}\! _{\mu } \, ,\, \, \, \, \, \, \, \, \, \, Z_{k}^{a} \, \, = \, \,
(Y^{-1} )_{k}^{a} \, ,\, \, \, \, \, \, \, \, \, \, A^{m}\! _{\mu } \, \, =\, \, h^{a} _{\mu } \, A_{a}^{m} \, .}
\end{array}
\end{equation}
It is easy to verify with the help of \eqref{GrindEQ__3_5_} and \eqref{GrindEQ__3_6_} that formula
\eqref{GrindEQ__3_4_} is valid.

\noindent \textbf{\textit{Preposition 1.2.}} We represent the Noether identities \eqref{GrindEQ__2_7_} as the
system of differential equations for the unknown function  ${\cal  L}_{Q} $ , the Principle of the minimal gauge
interaction (Postulate 4) being taken into account. The solvability conditions of the second system of these
equations are satisfied if the unknown matrix functions  $Z$  and  $S$  have the form,
\begin{equation} \label{GrindEQ__3_7_}
\begin{array}{l}
{S_{ma}^{n\mu } \, \, =\, \, \delta _{m}^{n} \, h^{\mu }\! _{a} \, ,\, \, \, \, \, \, \, \, \, \, S_{ka}^{n\mu }
\, \, =\, \, 0\, ,\, \, \, \, \, \, \, \, \, \, S_{ma}^{l\mu } \, \, =\, \, 0\, ,\, \, \, \, \, \, \, \, \, \,
S_{ka}^{l\mu } \, \, =\, \, \delta _{k}^{l} \, h^{\mu }\! _{a} \, ,} \\ {Z\, \, =\, \, {\rm det}\, {\rm
(}Z_{k}^{a} {\rm )}\, {\rm ,} \, \, \, \, \, \, \, \, \, \, h\, \, {\rm =}\, \, Z\, \hat{h}\, \, =\, \, {\rm
det}\, (Z_{k}^{a} {\rm )}\, {\rm det}\, {\rm (}\hat{h}^{a}\! _{\mu } )\, \, =\, \, {\rm det}\, {\rm (}h^{a}\!
_{\mu } )\, .}
\end{array}
\end{equation}

\noindent \textbf{\textit{Preposition 1.3}}.After substituting the results of Prepositions 1.1 and 1.2 into the
first system of the equations \eqref{GrindEQ__2_7_} we shall see that this system of equations is satisfied
identically by the Lagrangian density \eqref{GrindEQ__3_2_} provided that the unknown matrix function  $U$  has
the form,
\begin{equation} \label{GrindEQ__3_8_}
U_{ma}^{n} \, \, =\, \, c_{m}\!^{n}\!_{ q} \, A_{a}^{q} \, \, -\, \, I_{m}\!^{b}\!_{ a} \, A_{b}^{n} \, , \, \, \,
\, \, \, \, U_{ka}^{n} \, \, =\, \, 0\, ,\, \, \, \, \, \, \, U_{ma}^{k} \, \, =\, \, I_{m}\!^{k}\!_{ l} \,
A_{a}^{l} \, \, -\, \, I_{m}\!^{l}\!_{ a} \, A_{l}^{k} \, ,\, \, \, \, \, \, \, U_{ka}^{l} \, \, =\, \, -A_{a}^{n}
\, I_{n}\!^{l}\!_{ k} \, .
\end{equation}

\noindent \textbf{\textit{Corollary}}. After substituting \eqref{GrindEQ__3_8_} and the first line of
\eqref{GrindEQ__3_7_} into \eqref{GrindEQ__2_2_}, we get the transformational laws of the gauge fields  $A_{a}^{R}
\, \, =\, \, \{ A_{a}^{m} ,\, A_{a}^{k} \} \, $ ,
\begin{equation} \label{GrindEQ__3_9_}
\begin{array}{l}
{\delta A_{a}^{m} \, \, =\, \, \omega ^{n} (x)\, c_{n}\!^{m}\!_{ q} \, A_{a}^{q} \, \, -\, \, \omega ^{n} (x)\,
I_{n}\!^{b}\!_{ a} \, A_{b}^{m} \, \, +\, \, h^{\mu } _{a} \, \partial _{\mu } \omega ^{m} (x)\, \, =\,
\, D_{a} \, \omega ^{m} (x)\, \, -\, \, \omega ^{n} (x)\, I_{n}\!^{b}\!_{ a} \, A_{b}^{m} \, ,} \\
{\delta A_{a}^{k} \, \, =\, \, \omega ^{m} (x)\, (I_{m}\!^{k}\!_{ l} \, A_{a}^{l} \, \, -\, \, I_{m}\!^{l}\!_{ a}
\, A_{l}^{k} )\, \, -\, \, A_{a}^{m} \, I_{m}\!^{k}\!_{ l} \, a^{l} \, \, h^{\mu } _{a} \, \, +\, \, h^{\mu } _{a}
\,
\partial _{\mu } a^{k} (x)\, \, } \\ {=\, \, D_{a} \, a^{k} (x)\, \, -\, \, \omega ^{m} (x)\, (I_{m}\!^{k}\!_{ l} \,
A_{a}^{l} \, \, -\, \, I_{m}\!^{l}\!_{ a} \, A_{l}^{k} )\, ,} \\
{\delta h^{\mu }\! _{a} \, \, =\, -\, \omega ^{m} (x)\, I_{m}\!^{a}\!_{ b} \, h^{\mu }\! _{b} \, \, +\, \, h^{\nu
}\! _{a} \, \partial _{\nu } \delta x^{\mu } \, ,\, \, \, \, \, \, \, \, \, \, \delta h^{a}\! _{\mu } \, \, =\, \,
\omega ^{m} (x)\, I_{m}\!^{a}\!_{ b} \, h^{b}\! _{\mu } \, \, -\, \, h^{a}\! _{\nu } \, \partial _{\mu } \delta
x^{\nu } \, ,\, } \\ {\delta A^{m}\! _{\mu } \, \, =\, \, D_{\mu } \, \omega ^{m} \, \, -\, \, A^{m}\! _{\nu } \,
\partial _{\mu } \delta x^{\nu } \, .}
\end{array}
\end{equation}
One of the main results of this corollary is that the tetrads  $h^{\mu }\! _{a} $  and  $h^{a}\! _{\mu } $  are
not the true gauge potentials in contrast to the usually accepted opinion.

The structure of the gauge field Lagrangian density is established by the following theorem.

\noindent \textbf{Theorem 2} (B.N. Frolov, 1999, 2003). The Lagrangian density
\begin{equation} \label{GrindEQ__3_10_}
{\cal  L}_{0} \, \, =\, \, h\, L_{0} (F^{m}\! _{ab} ,\, \, T^{c}\! _{ab} )\, ,\,
\end{equation}
where
\begin{equation} \label{GrindEQ__3_11_}
\begin{array}{l}
{F^{m}\! _{ab} \, \, =\, \, 2h^{\lambda }\! _{[a} \, \partial _{|\lambda |} A_{b}^{m} \, \, +\, \, C^{c}\! _{ab}
A_{c}^{m} \, \, -\, \, c_{n}\!^{m}\!_{ q} \, A_{a}^{n} A_{b}^{q} \, ,} \\
{T^{c}\! _{ab} \, \, =\, \, C^{c}\! _{ab} \, \, +\, \, 2I_{n}\!^{c}\!_{ [a} \, A_{b]}^{n} \, ,\, \, \, \, \, \, \,
\, \, \, C^{c}\! _{ab} \, \, = \, \, -2h^{c}\! _{\tau } \, h^{\lambda }\! _{[a} \partial _{|\lambda |} h^{\tau }\!
_{b]} \, \, =\, \, 2h^{\lambda }\! _{a} \, h^{\tau }\! _{b} \, \partial _{[\lambda } h^{c}\! _{\tau ]} \, ,}
\end{array}
\end{equation}
satisfies to the Principle of the local invariance (Postulate 1).

\textbf{}

\noindent \textbf{\large 4. Field equations of the gauge fields}
\setcounter{section}{4}\setcounter{equation}{0}

\textbf{}

\noindent The gauge field equations are the following,
\begin{equation} \label{GrindEQ__4_1_} \frac{\delta {\cal  L}_{0} }{\delta A_{a}^{k} } \, \, =\, \, -
\frac{\partial {\cal  L}_{Q} }{\partial A_{a}^{k} } \, ,\, \, \, \, \, \, \, \, \, \, \, \, \, \, \, \, \, \, \,
\frac{\delta {\cal  L}_{0} }{\delta A_{a}^{m} } \, \, =\, \, -\frac{\partial {\cal  L}_{Q} }{\partial A_{a}^{m} } \, .
\end{equation}
The right sides of these field equations can be represented in the form,
\begin{equation} \label{GrindEQ__4_2_}
\begin{array}{l}
{-\frac{\partial {\cal  L}_{Q} }{\partial A_{a}^{k} } \, =\, \, Z_{l}^{a} \, \left({\cal  L}_{Q} \, \delta
_{k}^{l}\, \, -\, \, \frac{\partial {\cal  L}_{Q} }{\partial P_{l} Q^{A} } \, P_{k} Q^{A} \right)\, \, =
\, \, ht^{a}\! _{k} \, ,\, } \\
{-\frac{\partial {\cal  L}_{Q} }{\partial A_{a}^{m} } \, \, =\, \, I_{m}\!^{l}\!_{b} \, x^{b} \, \left(ht^{k}\!
_{l} \right) \, \, +\, \, \frac{\partial {\cal  L}_{Q} }{\partial D_{a} Q^{A} } \, I_{m}\!^{A}\!_{B} \, Q^{A} \,
\, =\, \, h\left(\hat{M}^{k}\! _{m} \, \, +\, \, J^{k}\! _{m} \right)\, .}
\end{array}
\end{equation}

The consequence of \eqref{GrindEQ__4_1_} and \eqref{GrindEQ__4_2_} is the theorem.

\noindent \textbf{Theorem 3} (B.N. Frolov, 1963, 2003) (\textit{The theorem on the source of the gauge field}).
The source of the gauge field, introducing by the localized group  $\Gamma (x)$ , is the Noether current,
corresponding to the non-localized group  $\Gamma $ .

\textbf{}

\noindent \textbf{\large 5. Geometrical interpretation}
\setcounter{section}{5}\setcounter{equation}{0}

\textbf{}

\noindent In the geometrical interpretation of the theory the quantities  $h^{a}\! _{\mu } $  and  $A^{m}\! _{\mu
} $ becomes tetrad fields and a Lorenz connection, and the quantities,
\begin{equation} \label{GrindEQ__5_1_}
\begin{array}{l}
{F^{m}\! _{\mu \nu } \, \, =\, \, F^{m}\! _{ab} \, h^{a}\!_{\mu } \, h^{b}\! _{\nu } \, \, =\, \, 2\, \partial
_{[\mu }
A^{m}\! _{\nu ]} \, \, -\, \, c_{n}\!^{m}\!_{ q} \, A^{n}\! _{\mu } A^{q}\! _{\nu } \, ,} \\
{T^{c}\! _{\mu \nu } \, \, =\, \, T^{c}\! _{ab} \, h^{a}\! _{\mu } \, h^{b}\! _{\nu } \, \, =\, \, 2\, \partial
_{[\mu } h^{c}\! _{\nu ]} \, \, +\, \, 2I_{n}\!^{c}\!_{ a} \, h^{a}\! _{[\mu } \, A^{n}\! _{\nu ]} \, ,\, }
\end{array}
\end{equation}
become curvature and torsion tensors respectively.

The following theorem is valid.

\noindent \textbf{Theorem 4} (B.N. Frolov, 1999, 2003). The system of the gauge field equations
\eqref{GrindEQ__4_1_} derived by the variation with respect to the gauge fields  $\{ A^{k} _{a} \, ,\, \, A^{m}
_{a} \} $  is equivalent to the system of the field equations derived by the variation with respect to the fields
$\{ h^{a}\! _{\mu } \, ,\, \, A^{m}\! _{\mu } \} $ ,
\begin{equation} \label{GrindEQ__5_2_}
\frac{\delta {\cal  L}_{0} }{\delta h^{a}\! _{\mu } } \, \, =\, \, - \frac{\delta {\cal  L}_{Q} }{\delta h^{a}\!
_{\mu } } \, ,\, \, \, \, \, \, \, \, \, \, \, \, \, \, \, \, \, \, \, \frac{\delta {\cal  L}_{0} }{\delta A^{m}\!
_{\mu } } \, \, =\, \, -\frac{\delta {\cal  L}_{Q} }{\delta A^{m}\! _{\mu } } \, .
\end{equation}

The first of the gauge field equation \eqref{GrindEQ__5_2_} can be represented in the form,
\begin{equation} \label{GrindEQ__5_3_}
\begin{array}{l}
{\partial _{\nu } \frac{\partial {\cal  L}_{0} }{\partial T^{a}\! _{\nu \mu } } \, \, =\, \, \frac{1}{2}
h\left(t_{(0)a}^{\mu } \, \, +\, \, t_{(Q)a}^{\mu } \right)\, ,\, \, \, \, \, \, \, \, \, \, ht_{(Q)a}^{\mu } \,
\, =\, \, {\cal  L}_{Q} \, h^{\mu }\! _{a} \, \, -\, \, \frac{\partial {\cal  L}_{Q} }{\partial D_{\mu } Q^{A} }
\, D_{a} Q^{A} \, ,} \\ {ht_{(0)a}^{\mu } \, \, =\, \, {\cal  L}_{0} \, h^{\mu }\! _{a} \, \, -\, \, 2F^{m}\!
_{a\, \nu } \, \frac{\partial {\cal  L}_{0} }{\partial F^{m}\! _{\mu \nu } } \, \, -\, \, 2T^{c}\! _{a\, \nu } \,
\frac{\partial {\cal  L}_{0} }{\partial T^{c}\! _{\mu \nu } } \, \, +\, \, 2A^{m}\! _{\nu } \, I_{m}\!^{c}\!_{ a}
\, \frac{\partial {\cal  L}_{0} }{\partial T^{c}\! _{\mu \nu } } \, ,\, }
\end{array}
\end{equation}
and the second one can be represented in the form,
\begin{equation} \label{GrindEQ__5_4_}
\begin{array}{l}
{\partial _{\nu } \frac{\partial {\cal  L}_{0} }{\partial F^{m}\! _{\nu \mu } } \, \, =\, \, -\frac{1}{2} h
\left(J_{(0)m}^{\mu } \, \, +\, \, J_{(Q)m}^{\mu } \right)\, ,\, \, \, \, \, \, \, \, \, \, hJ_{(Q)m}^{\mu } \, \,
= \, \, \frac{\partial {\cal  L}_{Q} }{\partial A^{m}\! _{\mu } } \, \, =\, \, \frac{\partial {\cal
L}_{Q}}{\partial D_{\mu } Q^{A} } \, I_{m}\!^{A}\!_{B} \, Q^{B} \, ,} \\
{hJ_{(0)m}^{\mu } \, \, =\, \, \frac{\partial {\cal  L}_{0}} {\partial A^{m}\! _{\mu } } \, \, =\, \,
2\frac{\partial {\cal  L}_{0} }{\partial F^{n}\! _{\mu \nu } } \, c_{m}\!^{n}\!_{ q} \, A^{q}\! _{\nu } \, +\, \,
2\frac{\partial {\cal  L}_{0} }{\partial T^{c}\! _{\mu \nu } } \, I_{m}\!^{c}\!_{ a}\, h^{a}\!_{\nu } \, .\,}
\end{array}
\end{equation}

The field equations \eqref{GrindEQ__5_3_} and \eqref{GrindEQ__5_4_} yield the conservational laws for the
canonical energy-momentum tensor  $t_{(Q)a}^{\mu } $  of the external field added by the energy-momentum tensor
$t_{(0)a}^{\mu } $ of the free gauge field, and for the spin current  $J_{(Q)m}^{\mu } $  of the external field
added by the spin current  $J_{(0)m}^{\mu } $ of the free gauge field,
\[
\partial _{\nu } \left(ht_{(0)a}^{\mu } \, \, +\, \, ht_{(Q)a}^{\mu } \right)\, \, =\, \, 0\, ,\, \, \, \, \, \,
\, \, \, \, \, \, \, \, \partial _{\nu } \left(hJ_{(0)m}^{\mu } \, \, +\, \, hJ_{(Q)m}^{\mu } \right)\, \, =\, \, 0\, .
\]

The field equations \eqref{GrindEQ__5_3_} and \eqref{GrindEQ__5_4_} can be also represented in a geometrical form,
\begin{equation} \label{GrindEQ__5_5_}
\begin{array}{l}
{D_{\nu } \frac{\partial {\cal  L}_{0} }{\partial T^{a}\! _{\nu \mu } } \, \, +\, \, F^{m}\! _{a\nu } \,
\frac{\partial {\cal  L}_{0} }{\partial F^{m}\! _{\mu \nu } } \, \, +\, \, 2T^{c}\! _{a\nu } \, \frac{\partial
{\cal L}_{0} }{\partial T^{c}\! _{\mu \nu } } \, \, -\, \, \frac{1}{2} {\cal  L}_{0} \,
h^{\mu }\! _{a} \, \, =\, \, \frac{1}{2} \, ht_{(Q)a}^{\mu } \, ,} \\
{D_{\nu } \frac{\partial {\cal  L}_{0} } {\partial F^{m}\! _{\nu \mu } } \, \, =\, \, -\frac{1}{2} hJ_{(Q)m}^{\mu
} \, \, +\, \, \frac{\partial {\cal  L}_{0} } {\partial T^{b}\! _{\nu \mu } } \, I_{m}\!^{b}\!_{ a} \, h^{a}\!
_{\nu } \, .\, }
\end{array}
\end{equation}
If we have  ${\cal  L}_{0} \, \, =\, \, h\, L_{0} (F^{m} _{ab} )\, $ instead of \eqref{GrindEQ__3_10_}, then the
first field equation \eqref{GrindEQ__5_5_} is simplified and generalizes the Hilbert--Einstein equation to
arbitrary nonlinear Lagrangians,
\[
F^{m}\! _{a\, \nu } \, \frac{\partial {\cal  L}_{0} }{\partial F^{m}\! _{\mu \nu } } \, \, -\, \, \frac{1}{2}
{\cal L}_{0} \, h^{\mu }\! _{a} \, \, =\, \, \frac{1}{2} \, ht_{(Q)a}^{\mu } \, .
\]

\textbf{}

\noindent \textbf{\large 7. Conclusions}
\setcounter{section}{7}\setcounter{equation}{0}

\textbf{}

\noindent The main result of the Theorem on the source of the gauge field (Theorem 4) is that the sources of PGTG
are not only the energy-momentum and the spin momentum tensors as in the Einstein--Cartan theory [6,7], but also
the orbital angular momentum tensor. The gauge  $t$ - and  $r$ -fields are generated together by the
energy-momentum, angular momentum and spin momentum tensors [5].

Therefore in PGTG rotating masses (for instant, galaxies, stars and planets), and also polarized medium should
generate the  $r$ -field. In [8] the influence of the rotating Sun on the planets moving via the torsion
generating has been investigated. Also a gyroscope on the Earth should change its weight subject to changing the
direction of rotation because of the interaction with the rotating Earth. There exist some experimental evidences
of such effects. In [9--12] the change of the weight of rotating bodies or polarized medium have been observed
that can be explained as the result of the interaction of these bodied and medium with rotating Earth. To this
subject one also may refer the results of the N. Kozyrev's experiments with gyroscopes [13] and the mysterious
$J-M$  relation between angular moments and masses of all material bodies in our Metagalaxy [14,15].

In Russia in Scientific Institute of Cosmic System (NIIKS of M.V. Khrunichev GKNP Center) some hopeful results
have been received in experiments, in which the possibility has been demonstrated of using the decrease of the
weight of rotating mass for constructing an engine that could move a body without any contacts with other bodies
and without ejection of any reactive masses [16].

In case of General Relativity (GR) torsion vanishes and one gets only one field equation with the metric
energy-momentum tensor as the source. But nevertheless the effects of GR depend on both of the  $t$ - and  $r$ -
fields. In particular, the Lense--Thirring effect and the Kerr solution are induced by the  $r$ -field. The
well-known problem of the constructing of the external source for the Kerr metric [17,18] may has its solution in
considering the angular momentum of the external medium as the source.

In GR the coupling constants both of the  $t$ - and  $r$ - fields are equal to the Einstein gravitational
constant. But in PGGT this choice is not determined by the theory and the coupling constants of the  $t$ -field
and  $r$ -field have not to be equal to each other. In PGTG these constants can have different values, which
should be estimated only on the basis of the experimental data [5].

\textbf{}

\noindent \textbf{\large References}

\textbf{}

\noindent [1] Frolov B.N. \textit{Vestnik of Moscow University, Ser. Phys., Astron.} N6, 48--58 (1963) (in
Russian).

\noindent [2] Frolov B.N. In: \textit{Gravity, Particles and Space--time} (Ed. P. Pronin and G. Sardanashvily)
World Scientific: Singapore (1996) 113--144.

\noindent [3] Frolov B.N. Problems of the theory of gravity with quadratic Lagrangians into spaces with torsion
and nonmetricity. The Second PhD Thesis, Moscow: MSU (1999) 316 p. (in Russian).

\noindent [4] Frolov B.N. \textit{Poincar\'{e}--Gauged Theory of Gravity}, Moscow: ``Prometey'' Publishing House,
MSPU (2003) 160 p. (in Russian).

\noindent [5] Frolov B.N. In \textit{Abstracts of 11th Int. Conf.} ``\textit{Theoretical and Experimental Problems
of General Relativity and Gravitation'' }(1--7 July, Tomsk, Russia), Tomsk (2002) 44--45.

\noindent [6] Kibble T.W.B. \textit{J. Math. Phys.}  \textbf{2,} 212--221 (1961).

\noindent [7] Trautman A. \textit{Symposia Math}. \textbf{12}, 139--162 (1973).

\noindent [8] Tucker R. In Proceedings of ``\textit{Symmetries and Gravity in Field Theory''} (Workshop in honor
of Prof. J. A. de Azcarraga. June 9-11, 2003. Salamanca (Spain)). To be published.

\noindent [9] Hayasaka H. \textit{Technology Report Tohoku Univ.} \textbf{43}, N2 (1978).

\noindent [10] Hayasaka H., Takeuchi S. \textit{Phys . Rev. Lett.} (1989).

\noindent [11] Rostschin V.V., Godin S.M. \textit{JTF Letters}, \textbf{26}, Iss. 24, 70--75 (2000) (in Russian).

\noindent [12] Volkov Ju.V. \textit{Estestvennyje i Tehnicheskije nauki}, \textbf{N1}, 19 (2002) (in Russian).

\noindent [13] Kozyrev N.A. \textit{Selective Works}, Leningrad Univ. Publishing House: Leningrad, 1991.

\noindent [14] Muradian R.M. \textit{Astrophys. Space Science} \textbf{69}, 339 (1980).

\noindent [15] Sistero R.F. \textit{Astroph. Lett}. \textbf{23}, 235--237 (1983).

\noindent [16] Men'shikov V.A., Akimov A.F., Kochekan A.A., Svetlichnyj V.A. \textit{Engines without ejection of
reactive masses: prerequisites and results}. Publishing House of NIIKS: Moscow, 2003. 225 p. (in Russian).

\noindent [17] Lopez C.A. \textit{Phys. Rev.} \textbf{D30}, 313 (1984).

\noindent [18] Burinskii A.Ya. \textit{Gravit.\&Cosm.} \textbf{8}, N4(32), 261--271 (2002) (preprint
hep--th/0008129).

\end{document}